\documentclass[aps,twocolumn,amsthm,amssymb,showpacs,superscriptaddress,notitlepage,longbibliography]{revtex4-1}
\usepackage[utf8]{inputenc}
\usepackage[colorlinks=true,linkcolor=blue,anchorcolor=red,citecolor=blue, urlcolor=blue]{hyperref}
\usepackage{bm}
\usepackage{bbm}
\usepackage{braket}
\usepackage{array}
\usepackage{graphicx}
\usepackage{color}
\usepackage{amsmath}
\usepackage{mathtools}
\usepackage[linesnumbered,ruled]{algorithm2e}
\usepackage{soul}
\usepackage[caption=false]{subfig}
\usepackage{mathrsfs}
\usepackage{braket}
\usepackage{float}

\newcommand{\rs}{\text{rs}}

\SetKwFor{ForEach}{for each}{:}{}

\SetInd{1em}{1em}

\begin{document}

\title{A simple universal routing strategy for reducing the connectivity requirements of quantum LDPC codes}

\author{Guangqi Zhao}
\email{guangqi\_zhao@outlook.com}
\affiliation{Quantum Science Center of Guangdong-Hong Kong-Macao, Shenzhen 518045, China}
\affiliation{Centre for Engineered Quantum Systems, School of Physics,
University of Sydney, Sydney, NSW 2006, Australia}

\author{Fei Yan}
\affiliation{Beijing Key Laboratory of Fault-Tolerant Quantum Computing, Beijing Academy of Quantum Information Sciences, Beijing 100193, China}

\author{Xiaotong Ni}
\email{xiaotong.ni@gmail.com}
\affiliation{Quantum Science Center of Guangdong-Hong Kong-Macao, Shenzhen 518045, China}

\begin{abstract}
Quantum low-density parity-check codes reduce quantum error correction overhead but require dense, long-range connectivity that challenges hardware implementation, particularly for superconducting processors. We address this problem by demonstrating that long-range connections can be reduced at the cost of increased syndrome extraction circuit depth.
Our approach is based on the observation that X and Z ancilla qubits form short loops with data qubits—a property that holds for any quantum code. This enables implementing stabilizer measurement circuits by routing data qubit information through ancilla qubits when direct connections are unavailable.
For bivariate bicycle codes, we remove up to 50\% of long-range connections while approximately doubling the circuit depth, with the circuit-level distance remaining largely preserved.
This method can also be applied to surface codes, achieving the same hexagonal connectivity requirement as McEwen et al.~\cite{mcewenRelaxingHardwareRequirements2023}.
Our routing approach for designing syndrome extraction circuits is applicable to diverse quantum codes, offering a practical pathway toward their implementation on hardware with connectivity constraints.
\end{abstract}

\maketitle

\section{Introduction}

While quantum bits provide significant computational advantages over classical bits, they are inherently more susceptible to errors. Quantum error correction (QEC) is therefore essential for executing long quantum circuits and algorithms. Among various QEC schemes, the surface code~\cite{kitaev2003fault,bravyi1998quantum} has emerged as particularly promising due to its local connectivity requirements, with recent experiments demonstrating sub-threshold error rates on superconducting platforms~\cite{google2025nature}.
However, the surface code's very low rate imposes steep resource overheads. This has motivated interest in quantum LDPC (qLDPC) codes with higher encoding rates~\cite{freedman2002z2,evra2022decodable,kaufman2021new,hastings2021fiber,panteleev2021quantum,Breuckmann_2021,Panteleev.2021,leverrier2022quantum,dinur2022good}.

Significant effort has gone into constructing and optimizing qLDPC codes for experimental systems~\cite{Tremblay2022,bravyi2024bbcode,Xu2023,liangPlanarQuantumLowdensity2025,steffanTileCodesHighEfficiency2025,shaw2025lowering,geherDirectionalCodesNew2025}. This will be a long-lasting but worthwhile process, as current platforms have yet to achieve low logical error rates even with the surface code.
A central objective is to reduce the number of required long-range operations.
In mobile platforms such as neutral atoms, shuttling time can dominate the cost of fault-tolerant computation.
In solid-state platforms such as superconducting processors, each direct long-range gate requires a physical coupler\cite{wang2025bbexp, songRealizationHighFidelityPerfect2025,magnardMicrowaveQuantumLink2020,campagne-ibarcqDeterministicRemoteEntanglement2018, yanEntanglementPurificationProtection2022, rengerSuperconductingQubitResonatorQuantum2025}. Such gates are typically slower, and increased connectivity degrades fidelity and amplifies crosstalk (see Appendix~\ref{sec:appendix_crosstalk}).

Bivariate bicycle (BB) codes~\cite{bravyi2024bbcode} are a recently introduced family of qLDPC codes that combine strong performance with a pseudo-2D layout and commonly use weight-6 stabilizers. In the standard “toric” layout, each measurement qubit typically performs four local and two long-range gates to data qubits~\cite{bravyi2024bbcode}. To further reduce connectivity, “morphing circuits”~\cite{shaw2025lowering} lower the local degree from four to three but still require two long-range couplers, leaving the main hardware bottleneck.

In this work, we introduce a routing-based syndrome measurement approach that systematically reduces long-range connectivity for BB codes. Our method removes up to 50\% of long-range connections while largely preserving circuit-level distance, at the cost of roughly doubling circuit depth. Concretely, the BB requirement can be reduced to four local connections and one long-range connection per qubit. The same idea transfers to the surface code, achieving degree-3 connectivity with performance comparable to mid-cycle circuits~\cite{mcewenRelaxingHardwareRequirements2023}. These reductions can simplify recent qLDPC demonstrations~\cite{wang2025bbexp} and enable less demanding multilayer layouts~\cite{mathewsPlacingRoutingNonLocal2025}.

This manuscript is organized as follows. In Sec.~\ref{sec:background}, we discuss background information, including the structure of surface and BB codes, as well as previous work on reducing connectivity requirements in surface codes. Then, in Sec.~\ref{sec:routing_surface_code}, we demonstrate our routing approach for the surface code with reduced connectivity, specifically showing how to perform syndrome measurements for the surface code on a hexagonal lattice via routing. For performance evaluation, we conduct circuit-level simulations using Stim~\cite{gidney2021stim} with a detector error model~\cite{gidney2021stim,Higgott2025sparseblossom} that fully captures error mechanisms at the circuit level (See Appendix~\ref{sub:dem} for a short introduction to the detector error model). Results show that our circuit preserves the code distance, meaning the circuit-level distance matches the code distance. By applying this routing strategy, we demonstrate that up to 50\% of long-range couplers can be eliminated while largely maintaining circuit-level distances. Furthermore, the logical error rate of our circuit is comparable to that of conventional circuits, despite the reduced connectivity requirements. The code used to conduct the simulation is available in our code repository on GitHub~\cite{github_repo}.

\section{Background}
\label{sec:background}

\subsection{Surface code}

Surface codes~\cite{bravyi1998quantum,freedman2001projective} represent the most developed quantum error correction approach and potentially the most promising codes realizable in the near term~\cite{google2025nature,google2023surface}, due to their local two-dimensional structure. For surface codes on an $L \times L$ square lattice, characterized as $[[n, k, d]] = [[L^2, 1, L]]$ quantum codes, they saturate the theoretical trade-off bound for two-dimensional codes where $kd^2 \leq O(n)$~\cite{Bravyi2010tradeoff}.

\begin{figure}[ht!]
  \centering
  \includegraphics[width=0.35\textwidth]{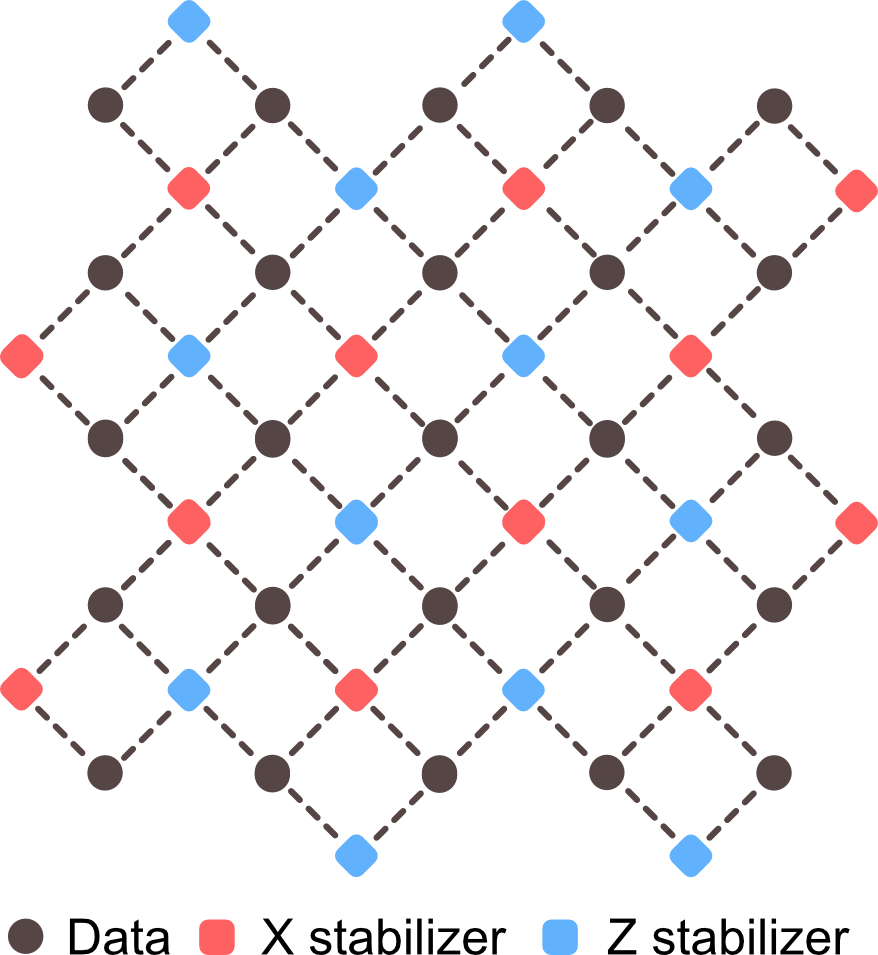}%
  \caption{2D square lattice layout of rotated surface code.}
  \label{figure:sc}
\end{figure}

Fig.~\ref{figure:sc} shows the structure of a rotated surface code. In the bulk, each ancilla qubit connects to four data qubits, and each data qubit connects to four ancilla qubits, resulting in 4-fold local connectivity in a 2D layout. Conventionally, syndrome extraction circuits for surface codes are implemented by performing CNOT gates across these connections.

\begin{figure}[ht!]
  \centering
    \subfloat[Z stabilizer measurement circuit]{%
        \includegraphics[width=0.28\textwidth]{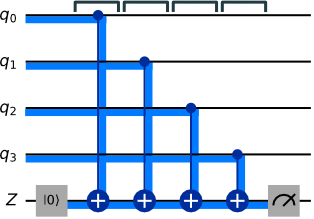}%
        \label{figure:sc_conventional_circuit_a}%
        } \\
    \subfloat[X stabilizer measurement circuit]{%
        \includegraphics[width=0.35\textwidth]{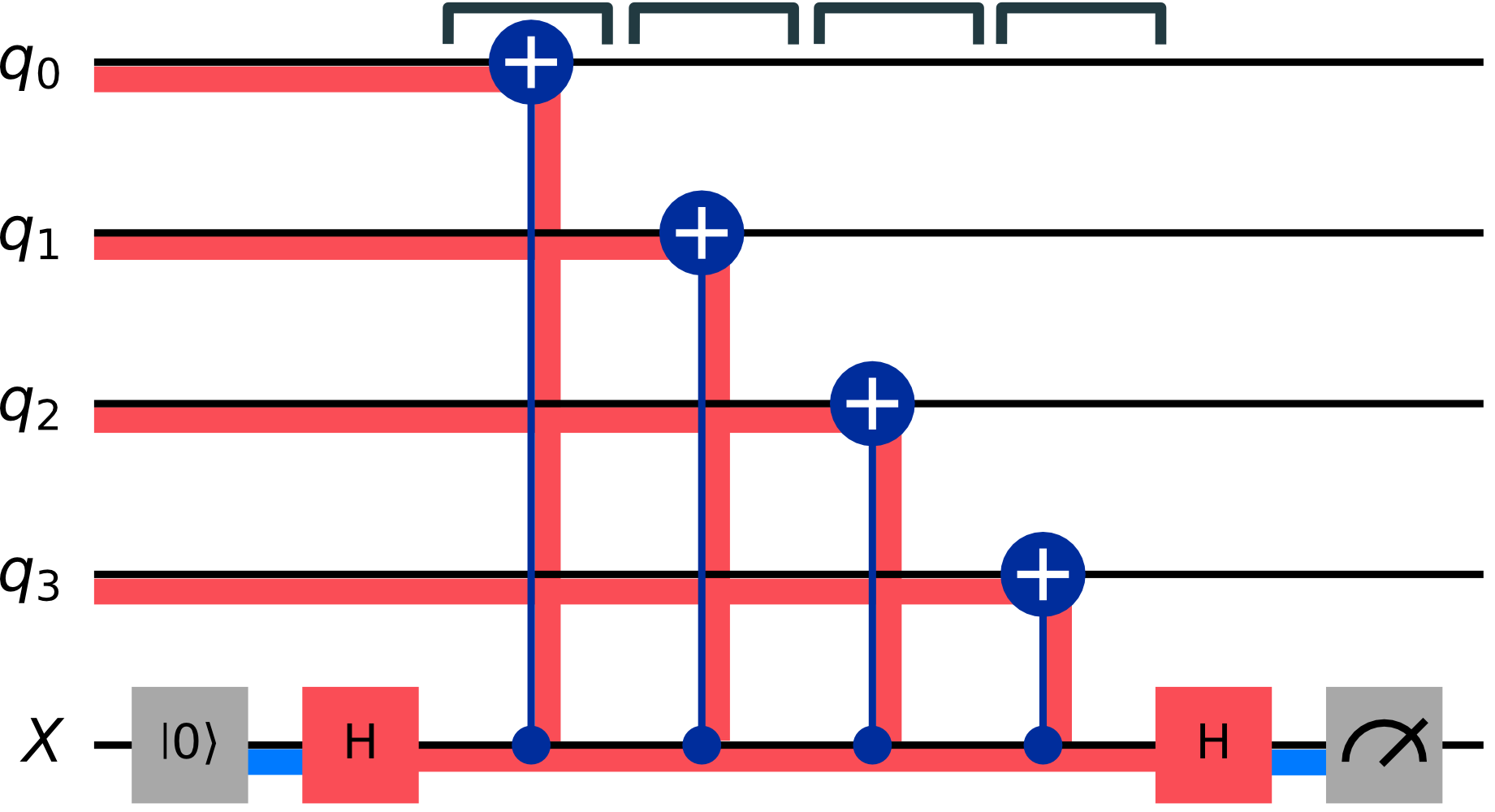}%
        \label{figure:sc_conventional_circuit_b}%
        }%
  \caption{Conventional syndrome extraction circuit for surface code. (a) shows the circuit for $Z$ stabilizer measurement, while (b) shows the $X$-type stabilizer measurement. Blue and red lines indicate the flows of stabilizer through the CNOTs.}
  \label{figure:sc_conventional_circuit}
\end{figure}

Although the conventional circuits for $X$ and $Z$ stabilizer measurements are well-known, for clear comparison with our new circuit discussed in the following sections, we illustrate them in Fig.~\ref{figure:sc_conventional_circuit} along with the backward transformation of the measured $Z$ stabilizers. We can see that we are indeed measuring the code stabilizer operators when no error occurs in the circuit. This is known as stabilizer flows, as described in~\cite{mcewenRelaxingHardwareRequirements2023}.

Through careful arrangement of operations, the circuit depth of surface codes can be optimized to just 4 layers of CNOT or CZ gates. This circuit has demonstrated impressive logical performance in experiments~\cite{Sebastian2022realizing,google2023surface,google2025nature,zhaoRealizationErrorCorrectingSurface2022,bluvstein_architectural_2025}, and it can be efficiently decoded through matching algorithms~\cite{dennis2002topological,fowler2013optimalcomplexitycorrectioncorrelated,higgott2022pymatching}.

\subsection{Bivariate bicycle codes}
\label{sec:bb_codes}
BB codes have recently attracted substantial interest due to their low overhead and high thresholds~\cite{bravyi2024bbcode}. These codes extend surface codes, as their stabilizers are translationally invariant and can be interpreted as multiple copies of the toric code~\cite{liangGeneralizedToricCodes2025}. In this section, we provide a brief introduction to BB codes.

Quantum Calderbank-Shor-Steane (CSS) codes~\cite{steane1996error,calderbank1996good} are a family of stabilizer codes~\cite{gottesman1997stabilizer} characterized by stabilizer generators that are exclusively $Z$-type or $X$-type operators. Similar to classical coding theory, these codes can be represented using parity check matrices $H_X$ and $H_Z$ for the $X$-type and $Z$-type stabilizers, respectively. The fundamental commutation requirements between $X$ and $Z$ stabilizers are elegantly captured by the matrix relation $H_X(H_Z)^T = 0$.

Generalized bicycle codes~\cite{kovalev2013,panteleev2021quantum} represent a class of CSS codes constructed using two commuting binary $\ell\times\ell$ matrices $A$ and $B$, where $AB = BA$. The parity check matrices for these codes are defined as:
\begin{eqnarray}
H_{\mathrm{X}}=[A, B] \text { and } H_{\mathrm{Z}}=\left[B^{\mathrm{T}}, A^{\mathrm{T}}\right] .
\end{eqnarray}
This construction naturally satisfies the commutativity condition, as $H_{\mathrm{X}} H_{\mathrm{Z}}^{\mathrm{T}}=A B+B A=\mathbf{0}$, thus yielding a valid CSS code. A particularly useful approach involves using binary circulant matrices for $A$ and $B$, since these matrices inherently commute~\cite{kovalev2013}. This formulation encompasses the original bicycle codes as a special case when $B=A^{\mathrm{T}}$~\cite{mackay2004sparse}.

BB codes represent a specialized class of generalized bicycle codes where matrices $A$ and $B$ are defined through polynomials $A = f(x,y)$ and $B = g(x,y)$. These polynomials belong to the Laurent polynomial ring $R := \mathbb{Z}_2[x,y,x^{-1},y^{-1}]$~\cite{haahCommutingPauliHamiltonians2013,liang2024extracting,liangGeneralizedToricCodes2025}. 

To align with conventional coding theory and parity check matrix formulations, we can represent the Laurent polynomial variables $x$ and $y$ in matrix form. Following Bravyi et al.~\cite{bravyi2024bbcode}, let $I_{\ell}$ and $S_{\ell}$ denote the identity matrix and the cyclic shift matrix of size $\ell \times \ell$, respectively. In the cyclic shift matrix $S_{\ell}$, the $i$-th row contains a single nonzero entry of one at column $i+1 \ (\bmod \ell)$. Then the variables $x$ and $y$ can be expressed as matrices:
\begin{eqnarray}
x=S_{\ell} \otimes I_m \quad \text { and } \quad y=I_{\ell} \otimes S_m
\end{eqnarray}

This representation has important properties: $xy=yx$ (commutativity) and $x^{\ell}=y^m=I_{\ell m}$ (periodicity). These properties ensure that polynomials in $x$ and $y$ maintain the necessary commutation relations when translated to the matrix formalism, facilitating the construction of valid quantum codes with predictable structures.

Consider the BB codes defined by a pair of matrices:
\begin{eqnarray}
A=A_1+A_2+A_3, \quad \text { and } \quad B=B_1+B_2+B_3,
\end{eqnarray}
where each component matrix $A_i$ and $B_j$ represents a power of $x$ or $y$. To prevent term cancellation, we require that the $A_i$ terms are distinct, and similarly, the $B_j$ terms are distinct. 

This construction ensures that matrices $A$ and $B$ have exactly three non-zero entries in each row and column. The commutativity requirement $AB=BA$ is automatically satisfied because $xy=yx$. Consequently, the BB code is LDPC, with parameters $[[n, k, d]]$ given by~\cite{bravyi2024bbcode}:
\begin{eqnarray}
n=2 \ell m, \quad k=2 \cdot \dim(\ker(A) \cap \ker(B)), \nonumber\\
d=\min \left\{|v|: v \in \ker\left(H_X\right) \backslash \rs\left(H_Z\right)\right\}.
\end{eqnarray}

Furthermore, to ensure that BB codes can be implemented with two-dimensional toric code layouts, matrices $A$ and $B$ must satisfy specific conditions outlined in Lemma 4 of Ref.~\cite{bravyi2024bbcode}.
For the examples used in this work, we can always write $A$ and $B$ in the form:
\begin{eqnarray}
    A=1+x+x^ay^b, \quad \text{and} \quad B=1+y+x^cy^d.
    \label{eq:bbcode_poly}
\end{eqnarray}
This construction provides a straightforward generalization of the standard toric code, and when $A = 1+x$ and $B=1+y$, this BB code reduces precisely to the conventional toric code. Thus, this family can be termed $(a,b,c,d)$-generalized toric codes~\cite{liangGeneralizedToricCodes2025}.

Similar to the surface code, we also need to put ancilla qubits in between data qubits to achieve  low-depth syndrome measurement circuits and to have 4 out of 6 couplers (2-qubit gates) to be nearest neighbor.
An example of such layout is shown in Fig.~\ref{figure:bb_code_long_range_connectivity}.

In this work, when we refer to connectivity, we specifically mean the connectivity between data and ancilla qubits.
For connectivity graphs such as Fig.~\ref{figure:bb_code_long_range_connectivity}, many length-4 loops are present. This is because $X$ and $Z$ stabilizers overlap on an even number of data qubits, creating these loops.
In the surface code layout shown in~\autoref{figure:sc}, all smallest plaquettes are such length-4 loops.
We illustrate these loops for a BB code example in Fig.~\ref{figure:bb_code_long_range_connectivity}.
The presence of these loops suggests that a certain number of couplers can be eliminated without compromising syndrome measurement capabilities.
Longer loops also exist, but we will not utilize them in this work.

\subsection{Previous works of reducing connectivity}

\label{sec:prev_works}

Recently, McEwen et al. demonstrated how to execute surface code with reduced connectivity requirements~\cite{mcewenRelaxingHardwareRequirements2023}.
They achieved implementation using only 3-fold connectivity in a 2D layout,  which forms a hexagonal lattice rather than a square lattice. 
We refer to this method as the mid-cycle approach, which is derived intuitively from the ``mid-cycle code'' of surface codes. This connectivity reduction comes at the cost of doubled circuit depth for 1 round of syndrome extraction.
The comparison of performance can be found in Fig.~\ref{figure:threshold_newcircuit_vs_4layer}. 
A similar approach is also explored for bivariate bicycle codes in~\cite{shaw2025lowering}, where the method is called morphing circuits.
The two main ideas of the mid-cycle and the morphing circuit approaches are
\begin{enumerate}
    \item Put the main focus on the code which is supported on both data and ancilla qubits. For example, this is the case when in the middle of a normal surface code measurement circuit.
    \item Using 2-qubit gates to transform a subset of stabilizers to single-qubit Pauli operators to measure them. Importantly, we only need partial connectivity between qubits in a stabilizer to achieve this. For example, 3 connections in a `C' shape is enough to shrink a weight-4 stabilizer.
\end{enumerate}
More discussion about the difference between the morphing circuit and our approach will be given in Sec.~\ref{sec:comparison_midcycle}.

A more straightforward and older approach to reducing connectivity is simply moving quantum information by using 2-qubit gates or LOCC teleportation~\cite{rosenbaum_optimal_2013}.
This is also related to the tasks such as quantum circuit mapping, which is about running quantum circuits on quantum processors with different connectivities. Compared to other approaches, circuit mapping in general does not change the QEC code used and does not consider circuit-level distance as its main objective.
Works in this direction include~\cite{lao_fault-tolerant_2020,li_tackling_2019, yin_qecc-synth_2025,berthusen2025toward}.

The problem of reducing connectivity is also closely related to the problem of handling processor defects in fault-tolerant circuits~\cite{augerFault2017,debroyLUCISurfaceCode2024}. There, the goal is to find an optimal circuit after certain qubits and connections are removed. Naturally, there are overlapping techniques in both of these two problems.

There are other methods for lowering connectivity requirements such as using subsystem codes~\cite{chamberland2020topo}, using floquet codes~\cite{hastings_dynamically_2021}, etc.

\section{Reducing connectivity through Routing: Surface code}
\label{sec:routing_surface_code}

In this work, we use $X$/$Z$-ancilla qubits for routing to reduce connectivity requirements. As a first step in designing such syndrome measurement circuits, we implement them to measure $X$ and $Z$ stabilizers sequentially. Specifically, the circuit follows the repetition pattern $C_X C_Z C_X C_Z \cdots$, where $C_X$ measures $X$ stabilizers and $C_Z$ measures $Z$ stabilizers. While it is generally not necessary to arrange the syndrome measurement circuits in this structure, designing a mixed $X$ and $Z$ stabilizer measurement circuit represents a challenging optimization problem~\cite{broshuis_small_2024}.

After choosing the circuit pattern $C_X C_Z C_X C_Z \cdots$, we observe that $X$ ancillae and their connections can be repurposed for measuring $Z$ stabilizers, and vice versa. This approach provides a straightforward method for reducing connectivity requirements across various quantum codes.

\begin{figure}[ht!]
  \centering
     \subfloat[rotated surface code with three couplers]{%
        \includegraphics[width=0.36\textwidth]{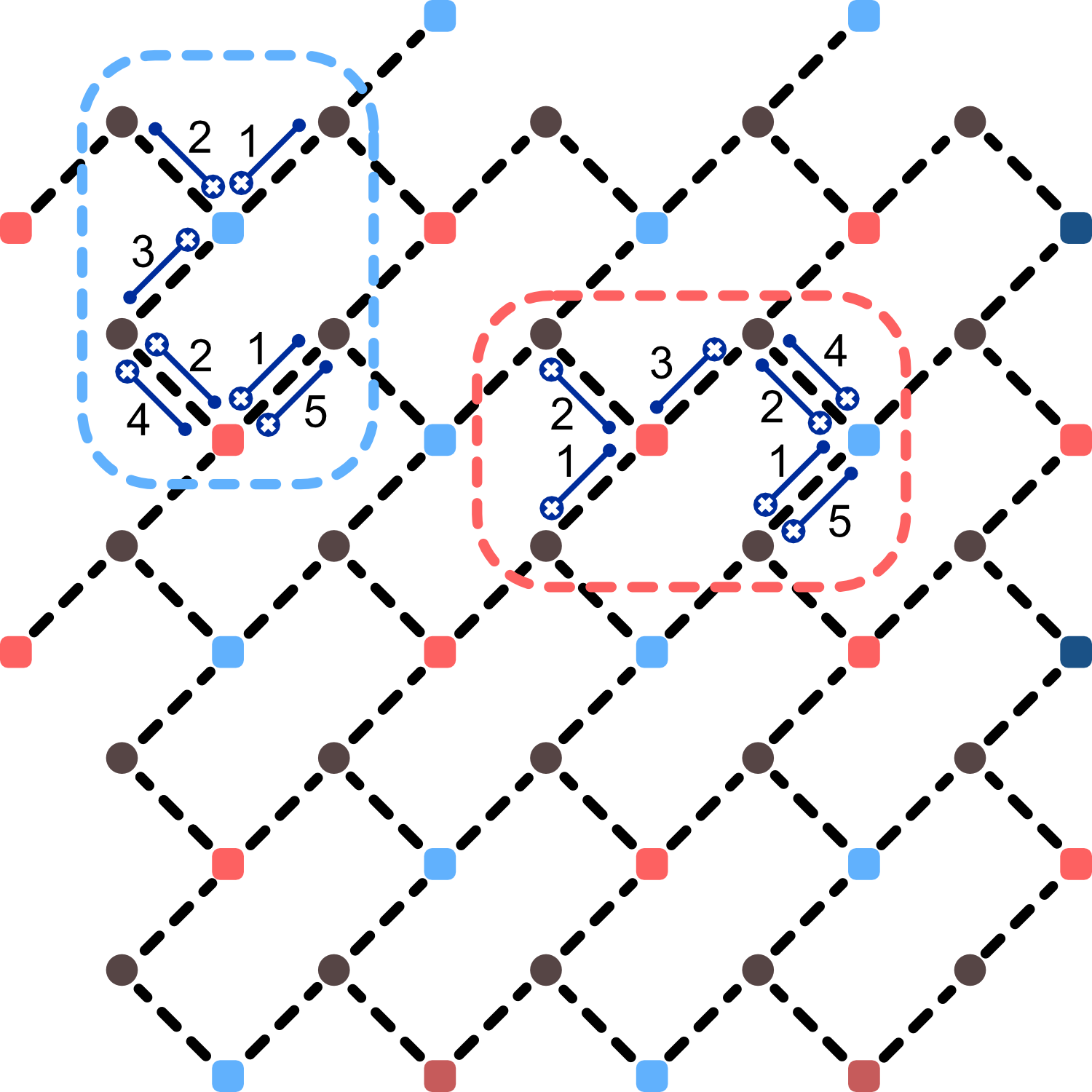}%
        \label{figure:sc_new_circuit_2d_a}%
        }%
        \quad
    \subfloat[flows of the $Z$ stabilizer through the circuit]{%
         \includegraphics[width=0.4\textwidth]{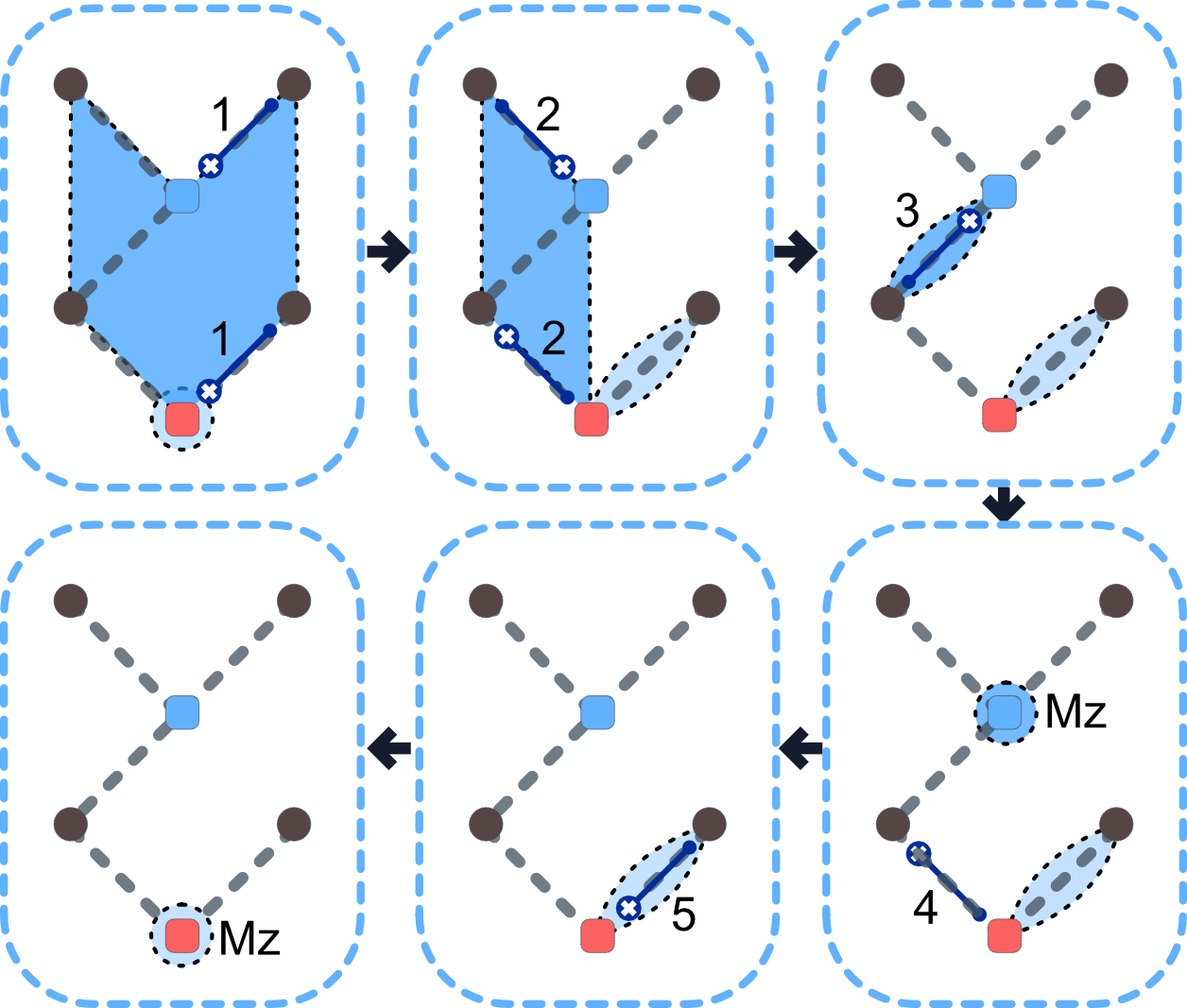}%
        \label{figure:sc_new_circuit_2d_b}%
        }%
  \caption{Stabilizer measurement is achieved by accessing adjacent ancilla qubits in this rotated surface code. (a) shows our arrangement allows for a 2D grid where all qubits (circles represent data qubits, and squares represent ancilla qubits) have at most three connections (dashed black lines), effectively forming a hexagonal lattice. In this configuration, stabilizer measurements utilize adjacent ancilla qubits of the opposite type; the blue dashed rectangle illustrates how a $Z$ stabilizer is measured using an $X$ ancilla located below it. Similarly, the red dashed rectangle demonstrates an $X$ stabilizer measurement via a $Z$ ancilla positioned to the right of it. On the lower and right boundaries, we place additional $X$ ancillas (indicated by dark red squares) and $Z$ ancillas (indicated by dark blue squares) to ensure all required measurements can be performed. The numbered CNOT gates indicate the sequential layers of operations required for syndrome extraction.
We show the evolution of two stabilizer operators in (b), one labeled with light blue and one with dark blue.
Note that the initial dark blue stabilizer operator is the 4-body surface code stabilizer multiplied by the $Z$ stabilizer of the ancilla qubit.}
  \label{figure:sc_new_circuit_2d}
\end{figure}

\begin{figure}[ht!]
  \centering
   \subfloat[Z stabilizer measurement via routing through a $X$ ancilla]{%
        \includegraphics[width=0.4\textwidth]{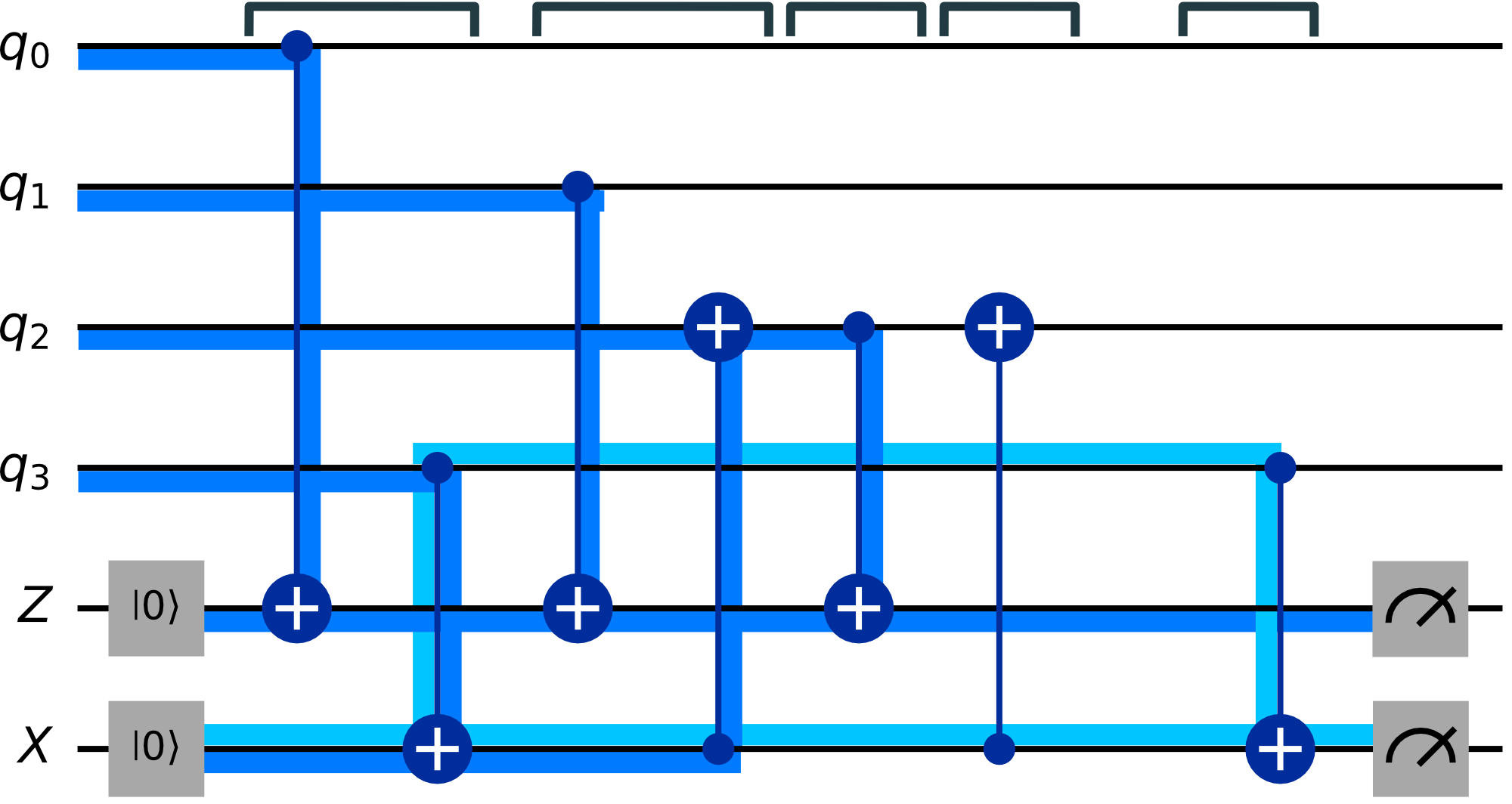}%
        \label{figure:sc_new_circuit_z}%
        }%
    \quad
    \subfloat[X stabilizer measurement via routing through a $Z$ ancilla]{%
         \includegraphics[width=0.44\textwidth]{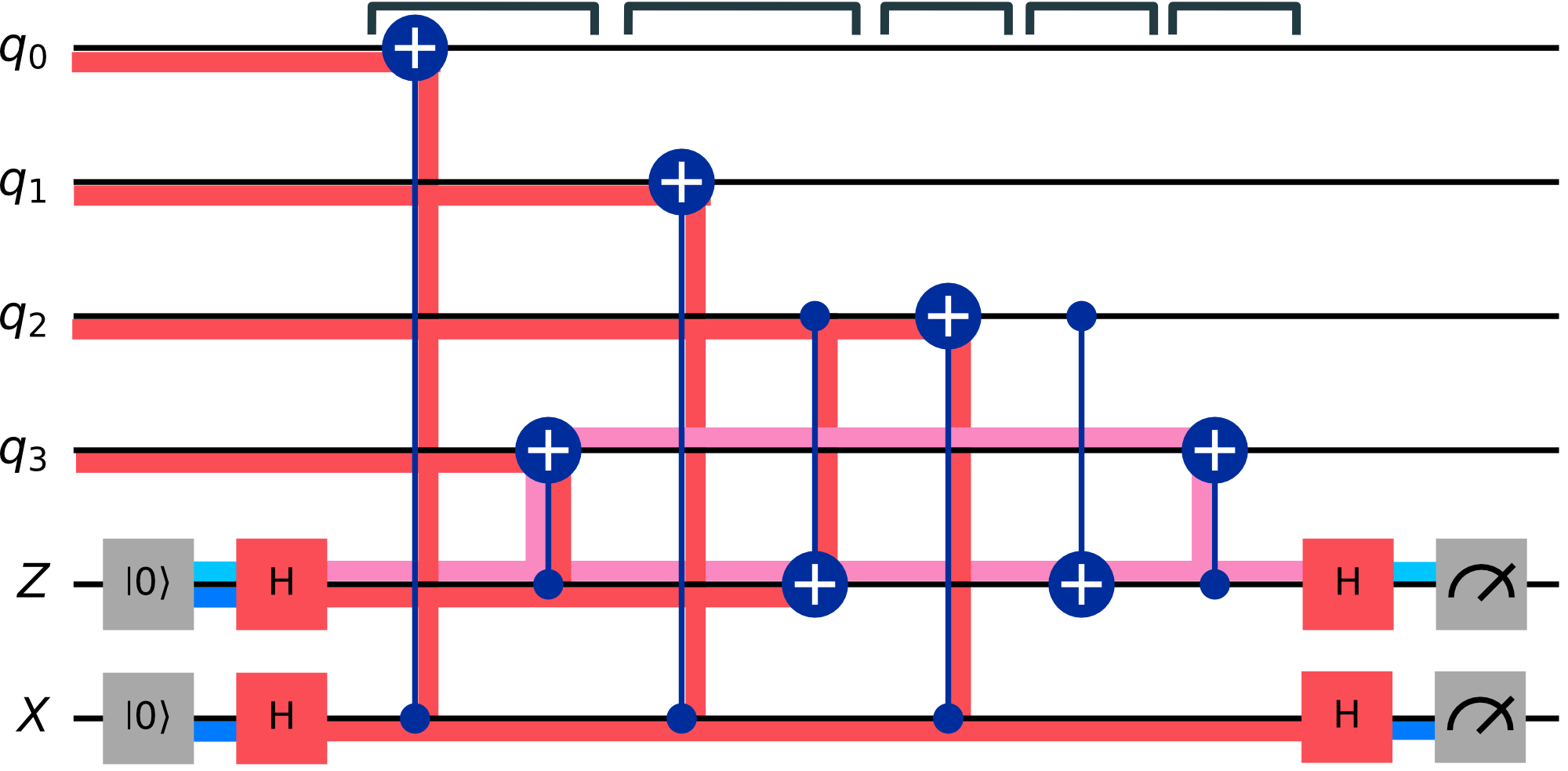}%
        \label{figure:sc_new_circuit_x}%
        }%
  \caption{Stabilizer extraction circuit for a single stabilizer in surface code with one coupler removed. Blue and red lines show the respective flows of stabilizer. (a) $Z$ stabilizer measurement when data qubit $q_3$ and $Z$ ancilla have no direct coupler; the information from $q_3$ is transferred through the $X$ ancilla via routing. (b) $X$ stabilizer measurement when data qubit $q_3$ and $X$ ancilla have no direct coupler; the information from $q_3$ is transferred through the $Z$ ancilla via routing. Note that the first three CNOT layers transfer stabilizer information to the $Z$ ancilla. The final two layers decouple data qubits from $X$ ancillae entangled during earlier operations.}
  \label{figure:sc_new_circuitx_and_z}
\end{figure}

Using this idea, we first implement a rotated surface code where each qubit has at most three connections. As shown in Fig.~\ref{figure:sc_new_circuit_2d_a}, the circuits within the blue and red dashed rectangles illustrate the measurement processes for $Z$ and $X$ stabilizers, respectively, with numbered CNOTs indicating the sequential order of CNOT gate operations. The first three layers of CNOTs transfer the parity of the four data qubits into the $Z$ ancilla. The final two CNOT layers reverse the routing and decouple the $X$ ancilla, which became entangled during the previous operations. To facilitate these detours, we add extra $X$ ancilla qubits at the lower boundary and extra $Z$ ancilla qubits at the right boundary, as shown in Fig.~\ref{figure:sc_new_circuit_2d_a}. A more traditional depiction of parts of the circuit is shown in Fig.~\ref{figure:sc_new_circuitx_and_z}.

\begin{figure}[htbp]
  \centering
  \includegraphics[width=0.49\textwidth]{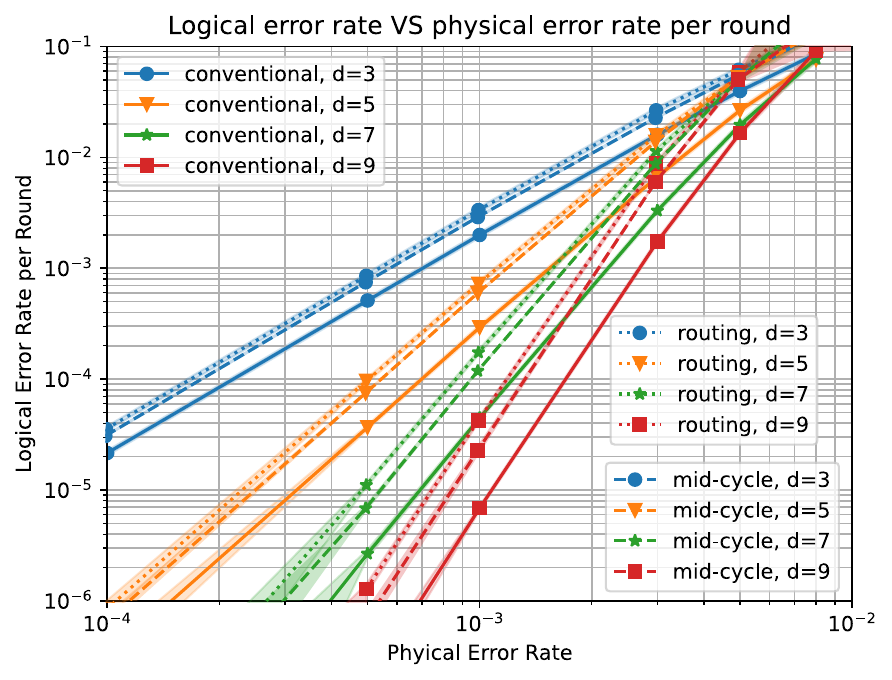}
  \caption{Physical error rate versus logical error rate per round under the SI1000 circuit-level error model for: conventional 4-layer surface code syndrome extraction circuit (solid line), mid-cycle three-coupler circuit (dashed line, generated with code provided by Ref.~\cite{mcewenRelaxingHardwareRequirements2023}), and our new three-coupler syndrome measurement circuit via routing (dotted line).
  We can see that while the mid-cycle approach has slightly better performance compared to the routing approach, the slopes of logical error rates are the same in the figure.
  }
  \label{figure:threshold_newcircuit_vs_4layer}
\end{figure}

One can check the effect of the $Z$ stabilizer measurement circuit on computational basis states. Because CNOT gates can be viewed as classical gates on such states, one can easily compute the combined transformation. Given four data qubits $\ket{x_1}, \ket{x_2}, \ket{x_3}, \ket{x_4}$, $\ket{0_z}$ as the $Z$ ancilla, and $\ket{0_x}$ as the $X$ ancilla, the CNOT gate transformation can be written as
\begin{equation}
    \ket{x_1 x_2 x_3 x_4 0_z 0_x}\rightarrow \ket{x_1 x_2 x_3 x_4 (x_1 \oplus x_2 \oplus x_3 \oplus x_4)_z 0_x}.
\end{equation}
Therefore, if no error occurs in this circuit, the $Z$ ancilla qubit measures the parity, and the $X$ ancilla qubit $\ket{0_x}$ remains unchanged.
The $X$ ancilla qubit $\ket{0_x}$ can thus be used as a flag qubit~\cite{chao_quantum_2018}.
However, we found that when using the MWPM decoder, including additional flag qubit information leads to worse performance (See Appendix~\ref{sub:results_diff_dem} for a detailed discussion).
Therefore, we do not use the flag qubits for the results shown in Fig.~\ref{figure:threshold_newcircuit_vs_4layer}.

The position of removed couplers and the sequence of CNOT gates must be carefully arranged to prevent hook errors~\cite{dennis2002topological}. The gate order can be selected similarly to conventional rotated surface code approaches, where data qubits involved in the first two CNOT layers form lines perpendicular to the direction of minimal $Z$-type logical operators~\cite{yu2014low,Yoder2017surfacecodetwist}. We confirmed through circuit-level simulation that our implementation preserves the code distance, using the same method as in~\cite{bravyi2024bbcode}. Moreover, the slopes of the logical error rates in Fig.~\ref{figure:threshold_newcircuit_vs_4layer} also suggest that the circuit-level distances are preserved.

In the numerical simulation corresponding to Fig.~\ref{figure:threshold_newcircuit_vs_4layer}, we used the SI1000 error model~\cite{Gidney2022benchmarkingplanar}, which is inspired by superconducting qubit systems (see Appendix~\ref{appendix:circuit_level_error_model} for a brief introduction).
We first constructed the syndrome measurement circuit and then generated the detector error models using Stim~\cite{gidney2021stim}. The decompose errors parameter was set to true, instructing Stim to suggest a decomposition of hyperedges into edges for decoding with the PyMatching decoder library~\cite{Higgott2025sparseblossom}.

In our implementation, $X$ and $Z$ stabilizer measurements each require 5 layers of CNOT gates, resulting in a total of 10 CNOT layers to complete one round of stabilizer measurements.
In comparison, the mid-cycle approach uses 8 layers of CNOT gates. To determine the logical error rates of these circuits, we conducted circuit-level simulations and compared them to the conventional 4-layer circuit.

Specifically, we plotted logical error rate versus physical error rate for different code distances under the SI1000 noise model in Fig.~\ref{figure:threshold_newcircuit_vs_4layer}. As a summary for the SI1000 model, the error rates from largest to smallest are measurement, two-qubit gates, and single-qubit gates, with the measurement error rate being 50 times that of single-qubit gates. We observe that the routing approach exhibits slightly worse logical error rates compared to the mid-cycle approach, and both perform worse than the conventional approach. However, the slopes of the logical error rates are the same, as shown in the figure, suggesting that the circuit-level distances are comparable across these approaches. In particular, this indicates that the circuit-level distances are equal to the code distances $d$ for the routing approach.

Our routing approach for surface codes provides a more direct generalization pathway across various quantum codes. Graphically, our method can be understood as removing a single connection from a connection loop containing both $X$ and $Z$ ancillae, and rerouting the information between the resulting endpoints through the remaining connections. In subsequent sections, we demonstrate how to apply our method to BB codes to reduce long-range connection requirements.

\section{Reduce connectivity for Bivariate bicycle codes}
\label{sec:routing_bb}

BB codes with polynomials defined in Eq.~(\ref{eq:bbcode_poly}) maintain a toric code layout supplemented with long-range connections. Each stabilizer ancilla connects to six data qubits—four local and two nonlocal.
These long-range connections are characterized by the parameters $a$, $b$, $c$, and $d$.
This toric layout, including these connections, is translation-invariant.

\begin{figure*}[ht!]
  \centering
  \includegraphics[width=0.7\textwidth]{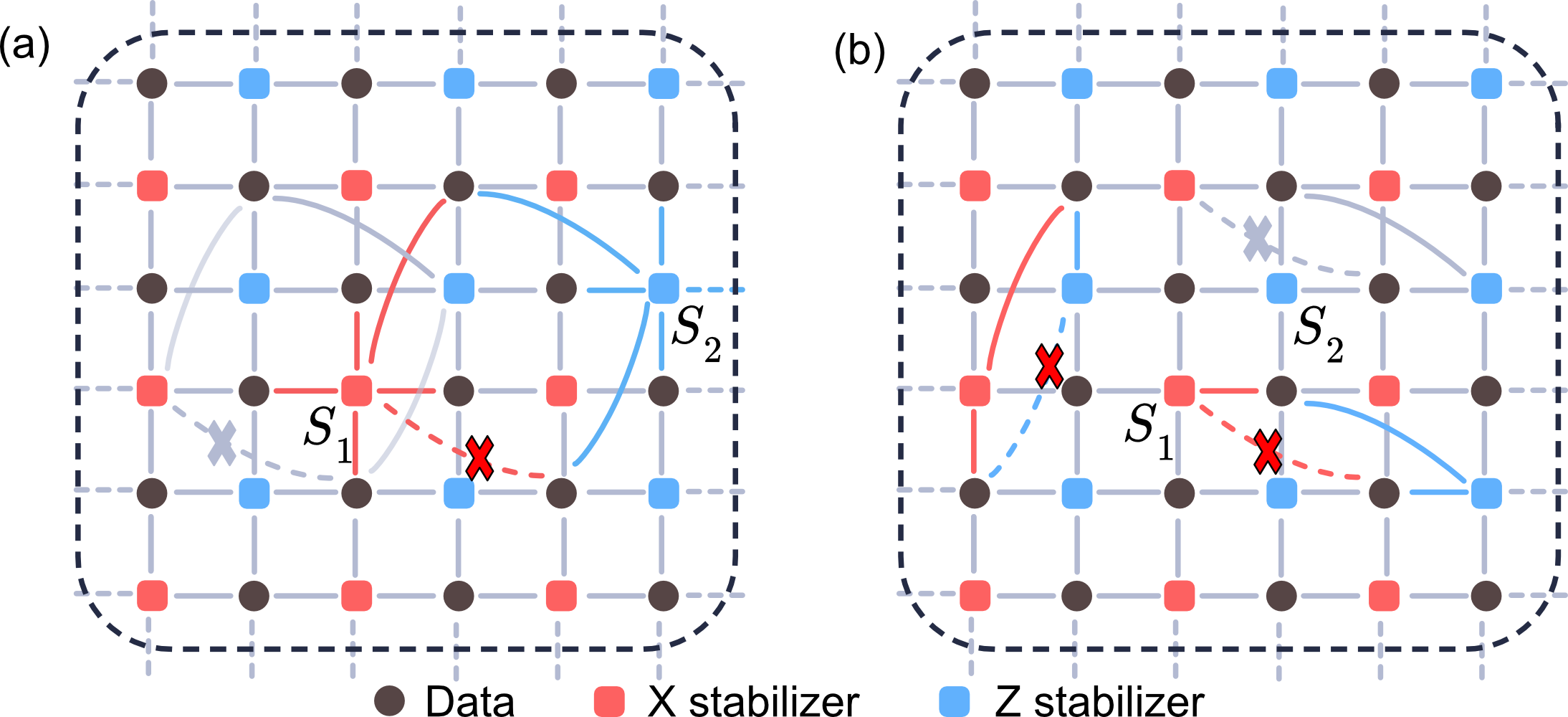}
  \caption{Torus layout of BB code with polynomials $A=1+x+xy$ and $B=1+y+xy$. (a) Each $Z$ ancilla pairs with an $X$ ancilla to form a length-four long-range loop with two data qubits. For these loops, our routing circuit enables removing one long-range connection per loop, reducing overall long-range connectivity by 25\%. (b) Alternative approach targeting short-long-short-long loops, where removing one long coupler per loop eliminates 50\% of all long-range couplers. This can also be viewed as removing an edge per $Z$ stabilizer in addition to the edges removed in (a). In both (a) and (b), gray loops indicate that there are many other loops we do not draw in the figure. Note that in this work, we remove couplers with translational symmetry for simplicity of circuit design. In general, there are other ways to remove couplers. However, this may come at the price of increased circuit depth.}
  \label{figure:bb_code_long_range_connectivity}
\end{figure*}

\begin{figure}[H]
  \centering
     \subfloat[Z stabilizer measurement via routing through a $X$ ancilla]{%
        \includegraphics[width=0.49\textwidth]{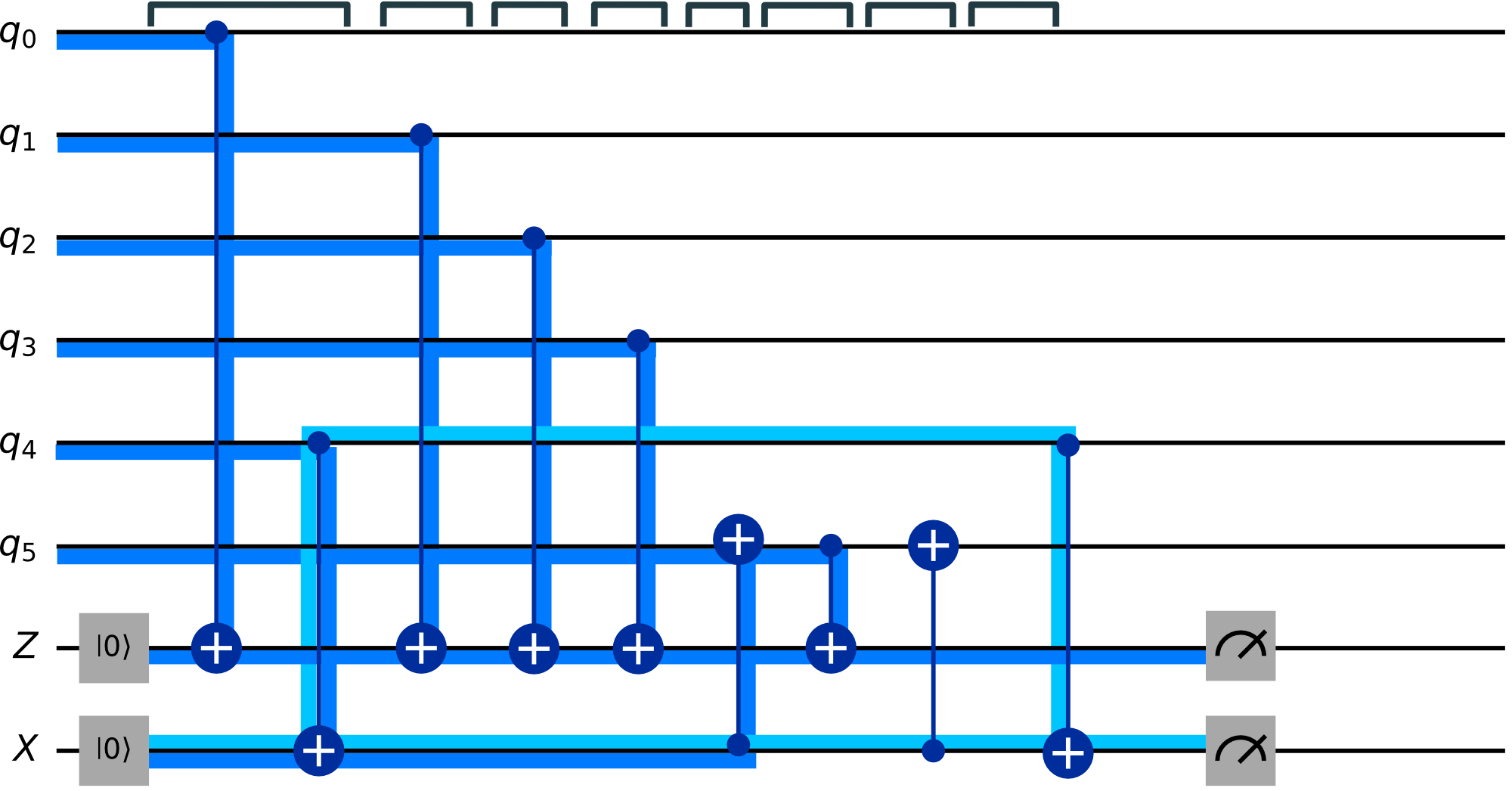}%
        \label{figure:bb_new_circuit_z}%
        }%
    \quad
    \subfloat[X stabilizer measurement via routing through a $Z$ ancilla]{%
         \includegraphics[width=0.49\textwidth]{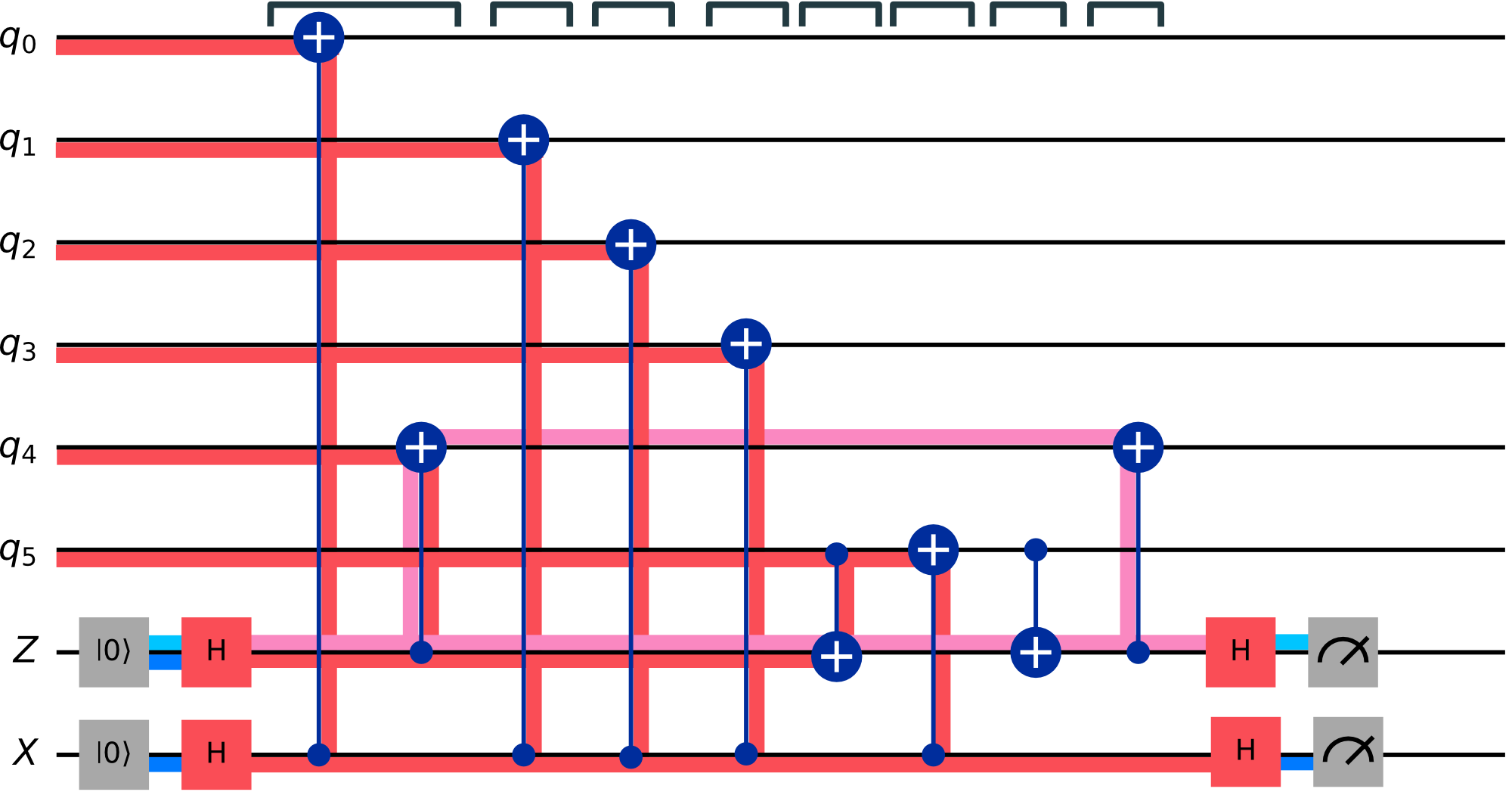}%
        \label{figure:bb_new_circuit_x}%
        }%
  \caption{Stabilizer extraction circuit for a single stabilizer in BB codes with one coupler removed. Blue and red lines show the respective flows of stabilizer. (a) $Z$ stabilizer measurement when data qubit $q_4$ and $Z$ ancilla have no direct coupler; the information from $q_4$ is transferred through the $X$ ancilla via routing. (b) $X$ stabilizer measurement when data qubit $q_4$ and $X$ ancilla have no direct coupler; the information from $q_4$ is transferred through the $Z$ ancilla via routing. Both circuits have 8 CNOT layers. Both circuits are designed to measure all $Z$ (or X) stabilizers simultaneous.  If measuring only a single stabilizer, both circuits could be optimized to 7 CNOT layers.}
  \label{figure:bb_code_circuit}
\end{figure}

Fig.~\ref{figure:bb_code_long_range_connectivity} illustrates an example configuration where $[a,b,c,d] = [1,1,1,1]$; a notable structural feature is that pairs of $X$ and $Z$ ancillae form length-four coupler loops with two data qubits. The existence of such loops allows us to apply the same connectivity reduction technique used in surface codes to BB codes. For each length-four long-range loop, we can remove one connection. Additionally, the translational symmetry of these length-four loops facilitates the design of circuits with lower depth. Following the strategy we used for the surface code, we measure $X$ and $Z$ stabilizers sequentially. As illustrated in Fig.~\ref{figure:bb_code_circuit}, both $X$ and $Z$ stabilizer measurements require a circuit of 8 CNOT layers if one coupler is removed.

There are different ways to remove connections.
For example, the two choices illustrated in Fig.~\ref{figure:bb_code_long_range_connectivity} eliminate 25\% and 50\% of all long-range connections, respectively.
These different choices result in slightly different depths for syndrome extraction circuits and logical error rates.
We perform simulations for both removal approaches: the ``three-quarters LR circuit", which retains 75\% of long-range (LR) couplers, and the ``half LR circuit", which retains 50\% of long-range couplers. 
In the three-quarters LR circuit, all $X$ stabilizers maintain full connectivity, while each $Z$ stabilizer has one long-range coupler removed. Specifically, we remove the same long-range coupler (the term $x^{-a}y^{-b}$ in polynomial $A^{T}$) from all $Z$ stabilizers. This configuration results in 6 CNOT layers (without routing) for all $X$ stabilizer measurements and 8 CNOT layers for all $Z$ stabilizer measurements.
In the half LR circuit, we remove the same long-range coupler from all $Z$ stabilizers (the term $x^{-a}y^{-b}$ in polynomial $A^T$) and the same coupler from all $X$ stabilizers (the term $x^cy^d$ in polynomial $B$).
The half LR circuit can also be viewed as removing an additional edge per $Z$ stabilizer compared to the three-quarters LR circuit.
This represents a general strategy that can be explored further in future work.
As long as short loops remain in the connectivity graph, we can continue to remove more edges.

We benchmark the performance of these two new circuits alongside IBM's original circuit~\cite{bravyi2024bbcode}.
The numerical simulations follow a similar approach to those used for the surface code, with the main difference being the choice of decoder. We construct the circuit and generate the detector error model using Stim with default settings.
First, we estimate the circuit-level distances, which serve as a useful initial metric for evaluating performance.
We employ the same method developed in~\cite{bravyi2024bbcode}.
Specifically, we randomly sample approximately 1000 different logical qubit errors with a trivial syndrome and use the decoder to find corresponding errors with small weights.
The decoder is a belief propagation decoder with ordered statistics (BP-OSD)~\cite{Panteleev2021degeneratequantum,roffe_decoding_2020,Roffe_LDPC_Python_tools_2022}.
We use the min-sum method for belief propagation and osd-0 for ordered statistics decoding.
The results are presented in Table~\ref{table:circuit_level_distance} for all three implementations, including the original length-7 circuits from IBM~\cite{bravyi2024bbcode}. Notably, the circuit-level distances for our two new circuits are identical, despite their different reduced connectivity, and are comparable to those achieved by IBM's circuit.

\begin{table*}
\begin{tabular}{ |c|c|c|c|c|c|}
 \hline
 $[[n, k, d]]$ & $\ell, m$ & (a,b,c,d) & $d_{3/4}$ & $d_{1/2}$ & $d_{\text{original}}$ \\
 \hline
 \hline
 $[[72,12,6]]$   & 6,6 & (3,-1,-1,3) &  $\leq 5$   &  $\leq 5$ & $\leq 6$  \\
 \hline
 $[[90,8,10]]$   & 3,15 & (0,5,-1,3) & $\leq 6$   &  $\leq 6$ & $\leq 8$ \\
 \hline
 $[[98,6,12]]$   & 7,7 & (1,-3,-3,1) & $\leq 8$   &  $\leq 8$ & NA\\
 \hline
 $[[108,8,10]]$   & 9,6 & (3,-1,-1,3) & $\leq 8$   &  $\leq 8$ & $\leq 8$ \\
 \hline
 $[[144,12,12]]$   & 12,6 & (3,-1,-1,3) &  $\leq 9$   &  $\leq 9$ & $\leq 10$  \\
 \hline
\end{tabular}
\caption{Some examples of BB codes and their parameters. All codes have the same polynomial structure $A = 1+x+x^ay^b$ and $B = 1 + y + x^cy^d$ with $(a,b,c,d)$ listed in the table. Variables $x,y$ are defined as $x = S_\ell \otimes I_m$ and $y = I_\ell \otimes S_m$ with $\ell,m$ also listed in the table. All BB codes feature weight-6 stabilizers. We display the circuit-level distance under different implementation scenarios: $d_{3/4}$ represents the circuit-level distance of three-quarters LR circuit (14 layers of CNOTs); $d_{1/2}$ shows the circuit-level distance for the half LR circuit (16 layers of CNOTs); and $d_{\text{original}}$ represents the circuit-level distance from IBM's results (7 layers of CNOTs)~\cite{bravyi2024bbcode} .}
\label{table:circuit_level_distance}
\end{table*}

\begin{figure*}[ht!]
  \centering
       \subfloat[]{%
        \includegraphics[width=0.48\textwidth]{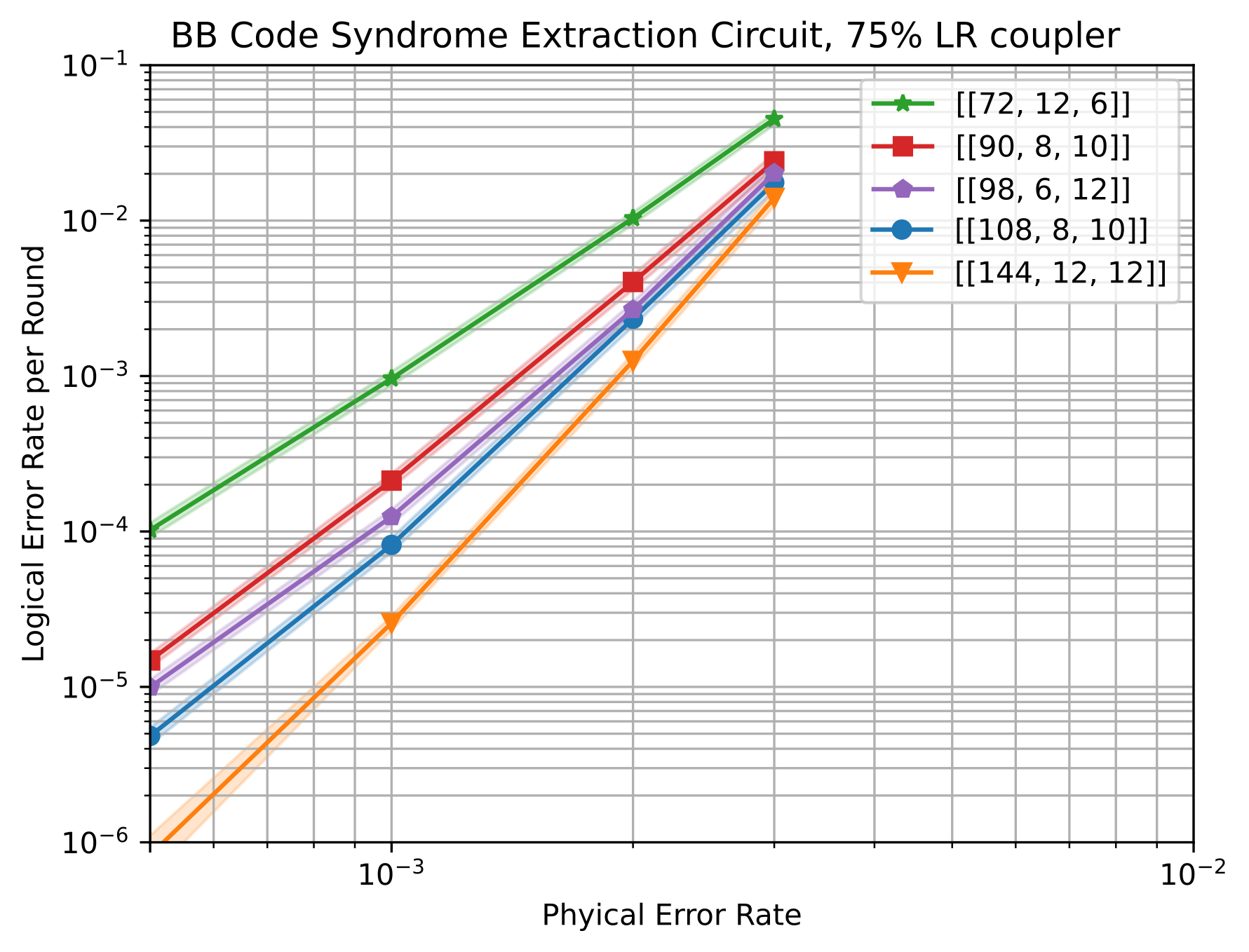}
        \label{figure:bb_code_threshold_a}%
        }%
    \quad
    \subfloat[]{%
         \includegraphics[width=0.48\textwidth]{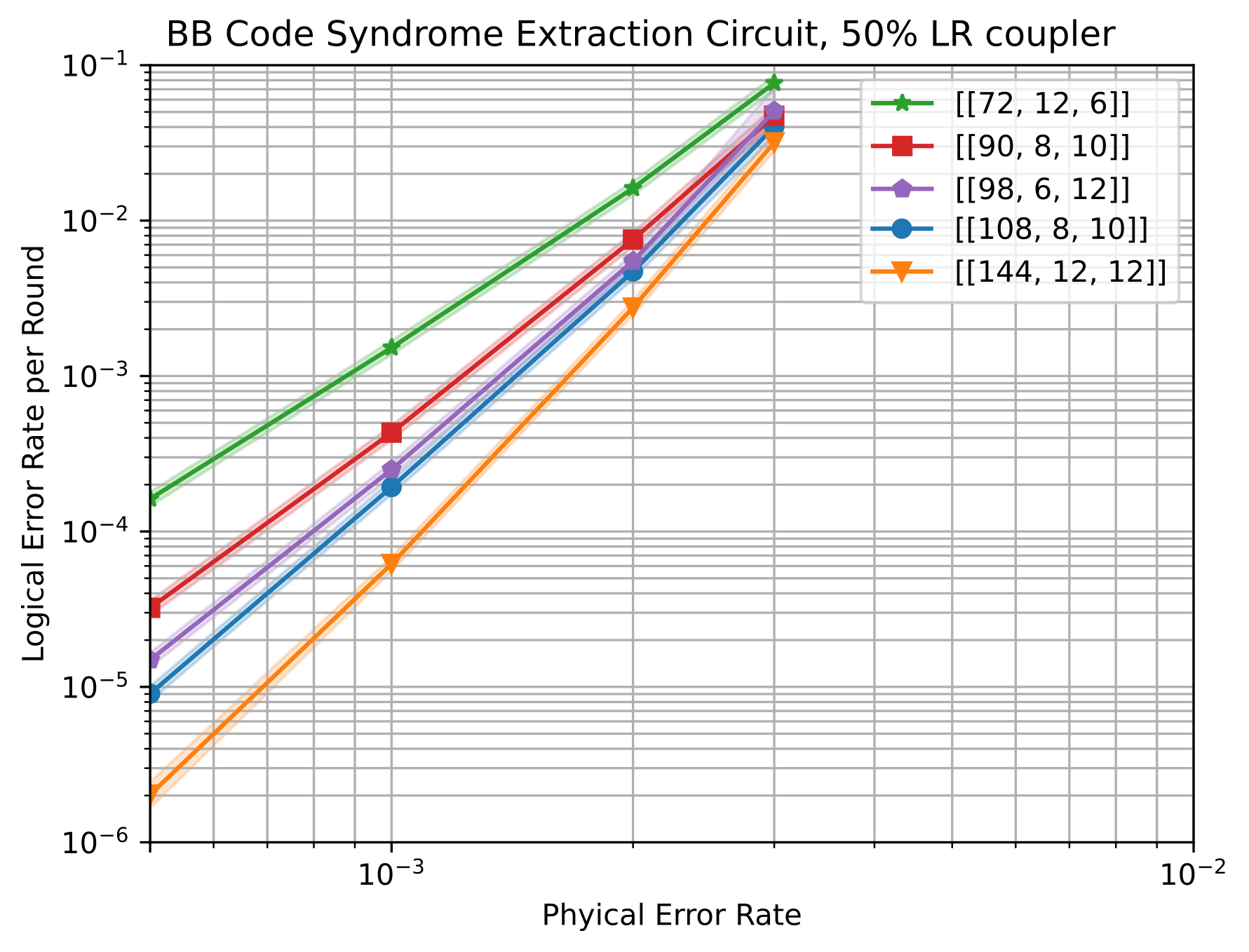}
        \label{figure:bb_code_threshold_b}%
        }%
  \caption{Total logical error rate per round versus physical error rate (of SI1000 model) for some BB codes. Numerical estimates of logical error rates were obtained by simulating d syndrome cycles for each distance-d code. (a) shows results for three-quarters LR circuit with 75\% of long-range (LR) couplers remain. (b) presents results for the half LR circuit with 50\% of long-range couplers remain. For the BP-OSD decoder, we used the 'osd-0' method and the min-sum algorithm.}
  \label{figure:bb_code_threshold}
\end{figure*}

\begin{figure}[ht!]
  \centering
    \includegraphics[width=0.48\textwidth]{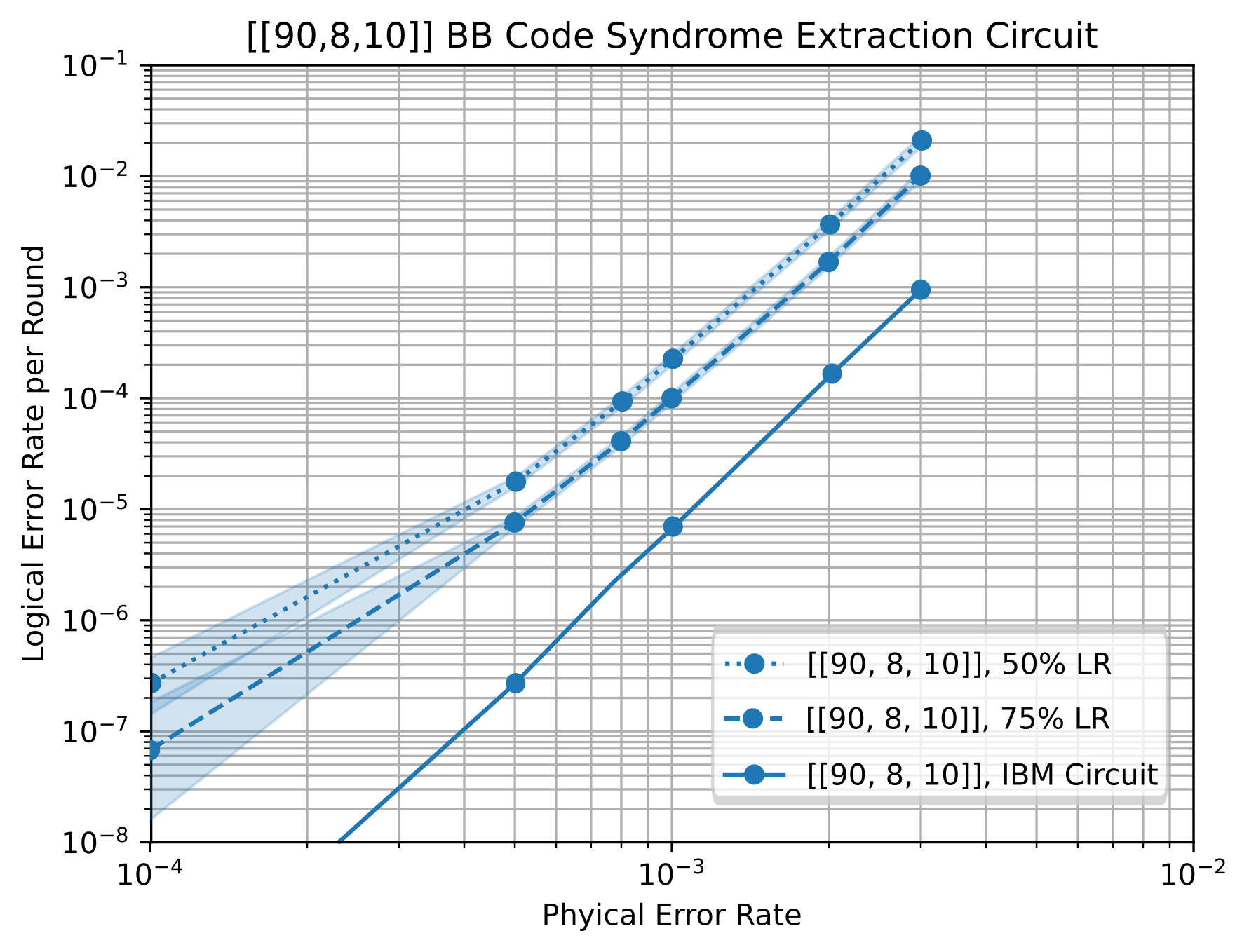}
  \caption{Logical versus physical error rate per round for [[90,8,10]] BB codes with different circuits: three-quarters LR circuit (75\% of long-range couplers retained, 14 layers of CNOTs), half LR circuit (50\% of long-range couplers retained, 16 layers of CNOTs), and IBM's circuit (points and lines obtained from Fig. 3 in Ref~\cite{bravyi2024bbcode}, 7 CNOT layers). Numerical estimates of logical error rates were obtained by simulating d syndrome cycles for each distance-d code. To facilitate comparison with IBM's results, we applied the standard depolarizing error model, where the error rates are the same value $p$ for all operations.}
  \label{figure:bb_routing_90}
\end{figure}

In Fig.~\ref{figure:bb_code_threshold}, we present logical error rates from circuit-level simulations for all codes listed in Table~\ref{table:circuit_level_distance}. These simulations were conducted using the SI1000 error model.
To compare the performance with IBM's simulation, which uses the standard depolarization error model~\cite{bravyi2024bbcode}, we simulate our circuits using the same model. Specifically, we compare our approaches with IBM's results for the [[90, 8, 10]] BB code, with results shown in Fig.~\ref{figure:bb_routing_90}.

At a physical error rate of $10^{-3}$, the logical error rates of our routing circuits are approximately ten times higher than those of IBM's implementation. This difference is expected, as our three-quarter LR routing circuit requires 14 CNOT layers and our half LR routing circuit needs 16 CNOT layers, whereas IBM's circuit uses only 7 CNOT layers. Despite this performance decline, our approach offers the significant advantage of eliminating up to 50\% of long-range couplers. The value of this tradeoff depends on future experimental capabilities; if implementing dense long-range connections proves more challenging than reducing physical error rates with fewer connections, then our scheme would be preferable.

\section{Compared to the morphing circuit approach}
\label{sec:comparison_midcycle}

As we mentioned in~\autoref{sec:prev_works}, the mid-cycle and morphing code approach comprises two components: one focuses on the code spanning both data and ancilla qubits, and the other involves choosing a circuit with low connectivity to contract stabilizers.
These two components are not fully dependent on each other.
The routing method used in this work solely addresses reducing the connectivity requirement.
It is likely that this routing method could be applied to the morphing circuit to further reduce connectivity, as the stabilizers still overlap on an even number of qubits.
However, this may require sacrificing circuit depth and potentially circuit-level distance.

In terms of logical error rates, we find that the morphing circuit approach performs better than the routing approach when using the same total number of qubits (with different codes). As a first step toward understanding this performance difference, we observe that the morphing circuit approach yields more syndrome data compared to the routing approach within the same amount of time. For the morphing circuit approach, switching between $[[2n,k,d]]$ and $[[n,k,\tilde{d}]]$, we obtain $2n$ bits of syndrome data with depth-14 circuits. However, for the routing approach with a code $[[n,k,\tilde{d}]]$, we only obtain $n$ bits of syndrome data with circuits of similar depth.
Further work is needed to gain more intuition about the performance of different syndrome measurement circuits.

\section{Conclusion and discussion}

In this work, we demonstrated that the hardware requirements for implementing qLDPC codes, particularly the number of long-range couplers, can be relaxed through routing techniques. Our simulation results show that, although routing circuits increase circuit depth, the circuit-level distance is not significantly impacted, and logical error rates remain within acceptable limits.
Our work is compatible with other research~\cite{mathewsPlacingRoutingNonLocal2025, steffanTileCodesHighEfficiency2025, liangPlanarQuantumLowdensity2025} on implementing qLDPC codes on superconducting processors and other connectivity-restricted systems.

This represents a trade-off between circuit depth and coupler density, and the value of this trade-off will depend on future experimental capabilities. If implementing dense long-range connections proves more challenging than reducing physical error rates with fewer connections, our scheme would be preferable. Looking to the long term, correlated errors arising from extensive connectivity can emerge as the dominant error source~\cite{ni_superconducting_nodate}, in which case our reduced-connectivity scheme would become increasingly advantageous.

Several questions remain open for future investigation. For example, the “half LR circuit” still exhibits redundancy in its couplers, suggesting further tradeoffs between circuit depth and connectivity for BB codes that warrant study. Future work could also establish methods applicable to general LDPC codes and develop optimization algorithms to identify minimal-depth circuits for systems with specific connectivity constraints.

We note that Zhou et al. have recently published work on a related topic~\cite{zhouLouvreRelaxingHardware2025}. However, our study was conducted independently and without prior knowledge of their research.

\emph{Acknowledgement}-We thank Craig Gidney for providing valuable discussion through the Quantum Computing Stack Exchange.
This work was supported by NSFC (No.92476206 and No.12322413) and Guangdong Provincial Quantum Science Strategic Initiative (GDZX2203001, GDZX2403001)

\bibliographystyle{apsrev4-1-etal-title}
\bibliography{ref}

\section{Appendix}
\label{sec:appendix}

\subsection{SI1000 error model}
\label{appendix:circuit_level_error_model}
SI1000 is a noise model with significantly different error rates for each operation, inspired by current superconducting processors~\cite{Gidney2022benchmarkingplanar}.
Let the probability $p$ represent the 2-qubit gate error rate. Single-qubit gates have a lower error rate of $p/10$. More importantly, the measurement error rate is $5p$, and because measurement duration is generally much longer than gate durations, idling qubits during measurement experience an error rate of $2p$. Therefore, this model heavily penalizes measurements, creating a larger gap between conventional and routed circuits compared to error models like the circuit depolarizing error model, where all operations have the same error rate. We note that larger measurement errors compared to gate errors are also observed in several other popular quantum computing experimental systems.

\subsection{Crosstalk in superconducting processor}
\label{sec:appendix_crosstalk}

More elements on a superconducting processor typically lead to increased crosstalk, from both classical control signals and qubit-qubit interactions.

When applying classical control signals to qubits, signal leakage often affects nearby qubits. As demonstrated in Ref.~\cite{Sebastian2022realizing}, this form of crosstalk can be mitigated through compensation signals. For operations such as simultaneous single-qubit gates, the control parameter space often exceeds the dimensionality of the single-qubit unitary group. By optimizing these control parameters during calibration, we can naturally mitigate most of this crosstalk.

More challenging is crosstalk arising from coupling between circuit elements, which manifests as unwanted interactions between qubits that decay with distance. Unlike control signal crosstalk, these interactions cannot be simply disabled when not performing two-qubit gates. Tunable couplers represent a common approach, allowing for effective cancellation of nearest-neighbor interactions by adjusting coupler qubit frequencies. However, this approach likely cannot eliminate all residual interactions, as documented in recent experiments~\cite{google2025nature,google2023surface, zajac_spectator_2021, harper_learning_2023} and theoretical studies~\cite{ni_superconducting_nodate}.

The widespread nature of the crosstalk problem is also evident in other symptoms.
The largest quantum error correction experiments on superconducting processors to date~\cite{google2025nature} highlight significant calibration challenges. As explained in Ref.~\cite{klimov_optimizing_2024}, optimal control parameters are not always found through optimization of local regions, indicating that complex residual interactions persist despite the use of tunable couplers.

To summarize, increasing the connectivity of superconducting processors presents two significant challenges: (1) designing hardware with consistently small unwanted couplings becomes increasingly difficult, and (2) calibration complexity grows substantially. For instance, frequency allocation—which must avoid similar frequencies for connected qubits—faces more constraints in qLDPC processors, leading to more complex optimization problems and longer calibration times. Therefore, trading increased circuit depth for reduced connectivity requirements becomes an attractive option when implementing quantum LDPC code circuits. This trade-off is particularly appealing as qubit coherence times improve, potentially offsetting the impact of longer circuit depths.

\subsection{Detector error model}
\label{sub:dem}

For current experimental systems, we often cannot directly measure multi-qubit stabilizers.
Even if some systems permit these operations, we still need to deal with the complex noise models associated with such operations. 
More commonly, a fault-tolerant circuit is decomposed into 2-qubit gates and other basic operations, and the circuit itself can be described by a classical LDPC codes~\cite{li2025prr}.
In the most basic quantum memory experiments, we still need to design and optimize syndrome measurement circuits using these operations.

In turn, the decoding task is about decoding the entire circuit rather than just the code itself. The detector error model~\cite{gidney2021stim,Higgott2025sparseblossom,derks2024design} provides one framework to describe how errors trigger detectors.
It is based on the fact that syndrome measurement circuits are Clifford circuits, which are efficiently simulatable. We can define detectors as parities of measurement outcome sets. These parities maintain fixed values when no errors occur in the circuit. Then by checking these parities, we can obtain information about where errors have occurred, allowing for further correction. A detector matrix $M$ offers a practical representation of these relationships. The kernel of this matrix encompasses all possible measurement outcomes that can occur in an error-free environment.

Given an error vector of whether errors happened in each location of the circuit, we first construct a measurement syndrome matrix $H$ that maps errors to their resulting syndromes. When combined with the detector matrix, we produce a comprehensive detector-error matrix $D = MH$. This integrated matrix plays a similar role as the parity checking matrix of a classical error correcting code.
The decoding can be done based on $D$.

In a quantum memory experiment, the circuit distance then can be defined as the minimum number of errors required to flip at least one logical operators of quantum code without triggering any detectors.

\subsection{Results with different detector error models and decoders}
\label{sub:results_diff_dem}

In our routing circuit, we measure $X$ and $Z$ stabilizers separately, yielding two measurement outcomes for each stabilizer measurement. For examples, as illustrated in Fig.~\ref{figure:sc_new_circuit_z} and Fig.~\ref{figure:bb_new_circuit_z}, $Z$ stabilizer measurement begins by resetting both $Z$ and $X$ ancilla qubits to the $Z$ basis. After applying the CNOT gates, we measure both ancillae in the $Z$ basis. The $Z$ ancilla measurement directly indicates the stabilizer value, as the circuit transfers stabilizer information to it. The $X$ ancilla measurement results could potentially provide additional information about error locations.

\begin{figure}[htbp]
  \centering
  \includegraphics[width=0.44\textwidth]{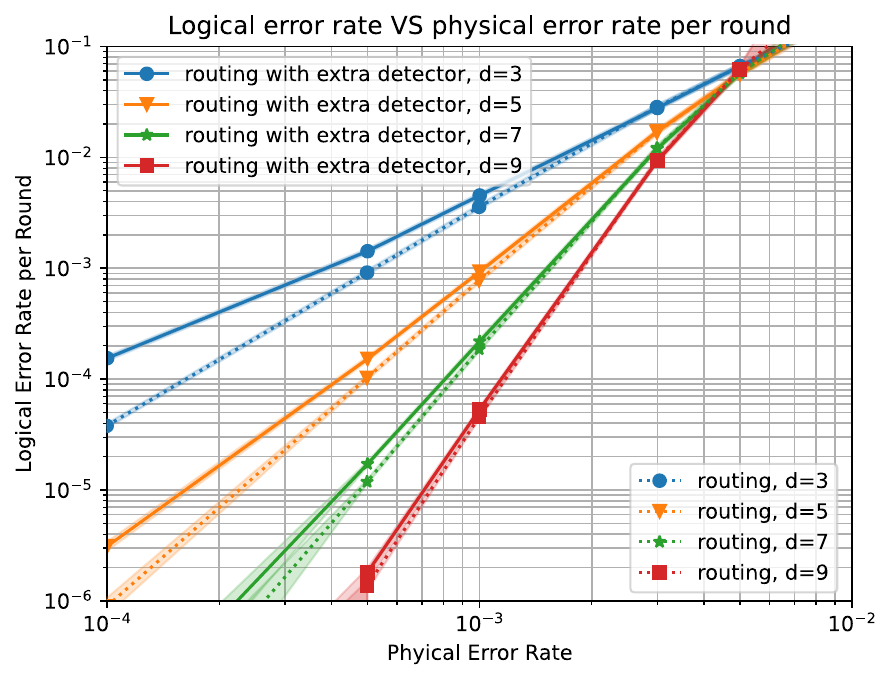}
  \caption{Logical versus physical error rate per round for surface code with different distances and detector settings (with MWPM decoder). Dashed lines show results with extra $X$ ancilla measurements (in $Z$ basis) as detectors, while solid lines show results without extra detectors. Notably, the configuration without extra detectors demonstrates better performance compared to the case with extra detectors. }
  \label{figure:sc_mwpm_diff_det}
\end{figure}

Intuitively, we would expect improved performance when incorporating $X$ ancillae measurement results (in $Z$ basis) into the detector error model. However, our simulations in Fig.~\ref{figure:sc_mwpm_diff_det} demonstrate this is not the case for MWPM decoder. Note that in our simulation, the detector error model is obtained with Stim. The decompose errors parameter is set to true.
This means that Stim will suggest a decomposition of hyper edges into edges so that the decoding can be performed with the PyMatching decoder library~\cite{Higgott2025sparseblossom}.
For more details about the whole simulation, please see the code at Github~\cite{github_repo}.

\begin{figure}[htbp]
  \centering
  \includegraphics[width=0.44\textwidth]{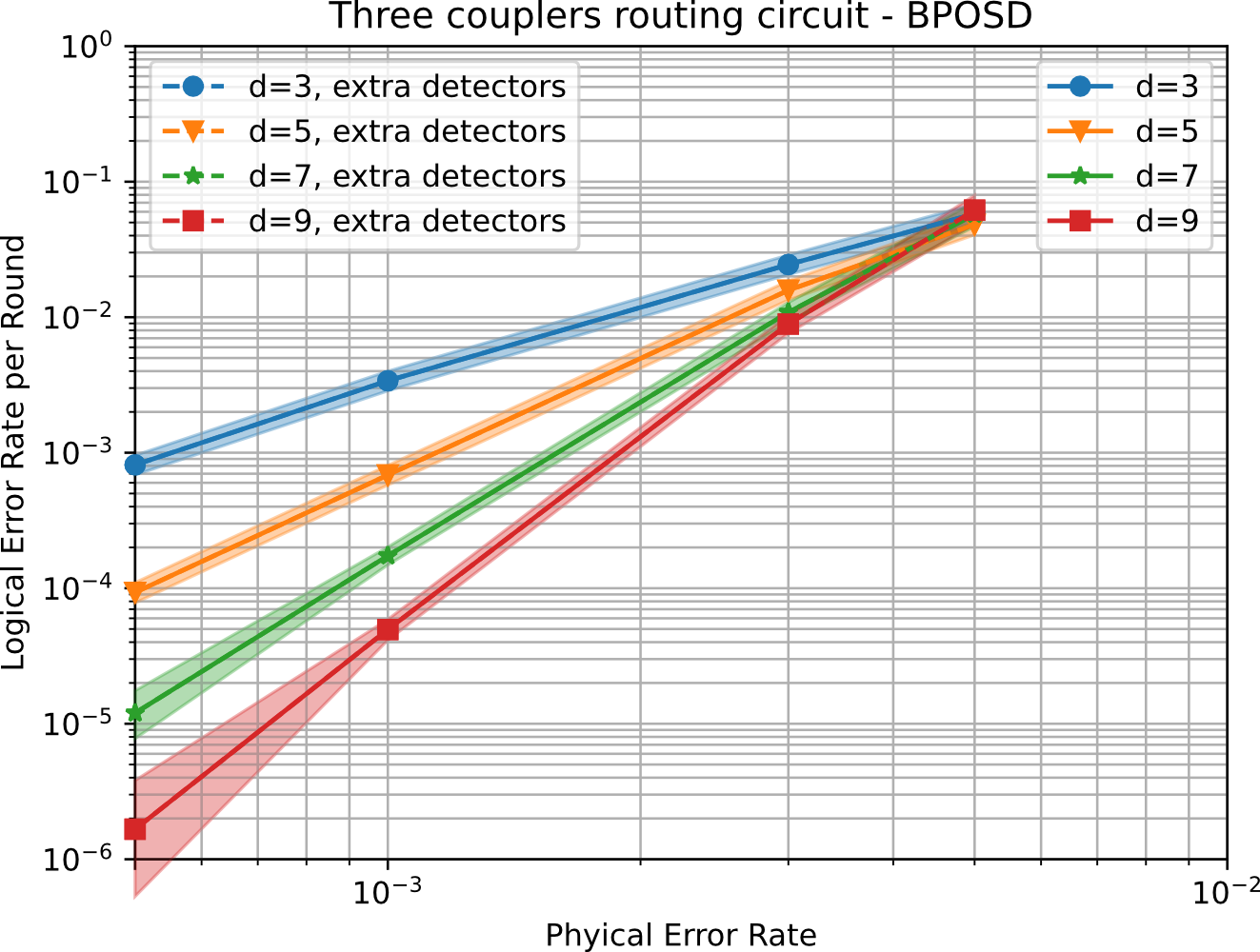}
  \caption{Logical versus physical error rate per round for surface code with different distances and detector settings (with BP-OSD decoder). Dashed lines show results with extra $X$ ancilla measurements (in $Z$ basis) as detectors, while solid lines show results without extra detectors. The difference between two detector error model is marginal (the dashed line lies under the solid line). In this simulation, we use 'osd-cs' method, and the osd order is set as 20. BP method is min-sum.}
  \label{figure:sc_BP-OSD_diff_det}
\end{figure}

It is natural to suspect the decoder might be responsible, as MWPM decoder is not necessarily good at handling hyper edges in the decoding graph. To investigate, we performed the same simulation using a BP-OSD decoder, with results shown in Fig.~\ref{figure:sc_BP-OSD_diff_det}. The results demonstrate that both detector error models performed almost identically under the BP-OSD decoder.
More investigations are needed to know the underlying reasons.

\end{document}